\documentclass{nature}
\usepackage{amssymb}
\usepackage{gensymb}
\usepackage{graphicx}

\title{Formation of the black-hole binary M33 X-7 via mass-exchange in a tight massive system}

\author{Francesca Valsecchi$^{1}$, Evert Glebbeek$^2$, Will M. Farr$^1$, Tassos Fragos$^{1, 5}$, Bart Willems$^1$, Jerome A. Orosz$^3$, 
 Jifeng Liu$^{4, 5}$ \& Vassiliki Kalogera $^1$}
\begin{document}
\maketitle
\begin{affiliations}
 \item Center for Interdisciplinary Exploration and Research in Astrophysics (CIERA) and Department of Physics and Astronomy, Northwestern University, 2145 Sheridan Road, Evanston, IL 60208, USA
 \item Department of Physics and Astronomy, McMaster University,  1280 Main Street West, Hamilton, Ontario, Canada L8S 4M1
 \item Department of Astronomy, San Diego State University, 5500 Campanile Drive, San Diego, CA 92182-1221, USA
 \item National Astronomical Observatories, Chinese Academy of Sciences, Beijing 100012, China
 \item Harvard-Smithsonian Center for Astrophysics, 60 Garden Street, Cambridge, MA 02138, USA
\end{affiliations}

\begin{abstract}
M33 X-7 is among the most massive X-Ray binary stellar systems known, hosting a rapidly spinning 15.65$\,$M$_\odot$ black hole orbiting an underluminous 70$\,$M$_\odot$ Main Sequence companion in a slightly eccentric 3.45$\,$day orbit\cite{2007Nature,Pietsch2006}. Although post-main-sequence mass transfer explains the masses and tight orbit\cite{Abubekerov2009}, it leaves unexplained the observed X-Ray luminosity, star's underluminosity, black hole's spin, and eccentricity.   
A common envelope phase\cite{2007Nature}, or rotational mixing\cite{DeMinkEtAl2009}, could explain the orbit, but the former would lead to a merger and the latter to an overluminous companion. A merger would also ensue if mass transfer to the black hole were invoked for its spin-up\cite{MorenoMendez2008}. Here we report that, if M33 X-7 started as a primary of 85-99$\,$M$_\odot$ and a secondary of 28-32$\,$M$_\odot$, in a 2.8-3.1$\,$day orbit, its observed properties can be consistently explained.
In this model, the Main Sequence primary transferred part of its envelope to the secondary and lost the rest in a wind; it ended its life as a $\simeq$16$\,$M$_\odot$ He star with a Fe-Ni core which collapsed to a black hole (with or without an accompanying supernova). The release of binding energy and, possibly, collapse asymmetries ``kicked'' the nascent black hole into an eccentric orbit. Wind accretion explains the X-Ray luminosity, while the black hole spin can be natal. 
\end{abstract}

M33 X-7 has been identified as an evolutionary challenge, given the massive components and its tight orbit relative to the large H-rich black hole (BH) companion. Four paths have been proposed to explain M33 X-7's formation, but none of them has addressed nor can simultaneously explain all its observed properties (Tables 1 and 2).
We performed detailed binary evolution calculations to explore possible evolutionary tracks. Given the spatial metallicity
 gradient of the M33 galaxy\cite{UetAl2009}, we assume a metallicity 50\% of the solar value for all our models.

To illustrate clearly M33 X-7's evolutionary history, we show the results for one of the successful evolutionary sequences in Figure 1. The progenitor comprises a primary of $\simeq$97$\,$M$_\odot$ (BH progenitor) and a secondary of $\simeq$32$\,$M$_\odot$ (BH-companion progenitor) in an orbit of $\simeq$2.9$\,$days.
During the first $\simeq$1.8$\,$ Myr the evolution is driven by mass loss via stellar winds, causing a decrease of the gravitational attraction between the components and expansion of the orbit to $\simeq$3.25$\,$days.
The more massive primary evolves faster than the secondary, growing in size to accommodate the energy 
 produced by fusing H into He at its center. Eventually, while still on its Main Sequence, 
 it expands and begins mass transfer (MT) onto the secondary when entering the sphere of influence of the secondary's gravitational field (through Roche-lobe overflow). 
 This stronger mode of mass loss brings the primary out of thermal equilibrium; in response the star shrinks, recovering its thermal equilibrium while always maintaining hydrostatic equilibrium, and hence dynamical stability.
During the first few tens of thousands of years of MT, the orbital period decreases  
because the more massive primary is transferring mass to the less massive secondary. 
When the secondary accretes enough matter to become the more massive component, the orbit starts expanding\cite{Verbunt1993}. 
The primary transfers most of its H-rich envelope and becomes a Wolf-Rayet star, and the strong Wolf-Rayet wind ($\sim$2 to 3$\cdot\rm{10}^{-5}$ $\,$M$_\odot$yr$^{-1}$) removes much of the remaining envelope, eventually interrupting the MT.
During the $\simeq$99,000$\,$ years of conservative MT, the original 32$\,$M$_\odot$ 
secondary becomes a massive $\simeq$69$\,$M$_\odot$ O-type star, while the primary becomes a $\simeq$51$\,$M$_\odot$ Wolf-Rayet. 	
Once the Wolf-Rayet wind sets in and the MT is interrupted, the wind blows away the remaining primary's envelope 
to expose the $\simeq$25$\,$M$_\odot$ He core. This mass loss drives further orbital expansion until the end of the primary's Main Sequence and throughout the core He burning phase.  At the same time, the now more massive secondary is losing mass via its own O-star wind at a lower rate 
 ($\sim10^{-6}\,$M$_\odot$yr$^{-1}$). 
At this time the orbit of the binary is circular and the spin period of each star is expected to be synchronized with the orbital period. 
The synchronization is due to exchange of angular momentum between the stars and their orbit caused by tidal interaction. 

The final stages of the primary's life  during and beyond carbon burning are too short ($\sim$60$\,$yr 
 for an initially $\simeq$25$\,$M$_\odot$ He star)\cite{TaurisHeuvel2003}
  to significantly change the stellar and orbital parameters. 
At the end of the primary's life, after $\simeq$3.7$\,$Myr,  M33 X-7 comprises a  $\simeq$16$\,$M$_\odot$ 
evolved Wolf-Rayet star with an Fe-Ni core, 
and a $\simeq$64.5$\,$M$_\odot$ O-star companion in a $\simeq$3.5$\,$day orbit.
Unable to support itself through further nuclear fusion, this massive He-rich star collapses into a BH, and a small fraction of the rest mass energy (10\%) is released as the BH's gravitational energy. Additionally, asymmetries in the
 collapse and associated neutrino emission $may$ impart an instantaneous linear momentum recoil (kick) to the newly born BH, even without any baryonic mass ejection at collapse.
Both these effects modify the orbital configuration inducing an eccentricity, and slightly decreasing the orbital separation to $\simeq$3.4$\,$days. In fact, while the release of binding energy leads to an increase of the orbital separation, the kick imparted to the BH acts to shrink it. For the remaining $\simeq$0.2$\,$Myr, the post-BH-formation binary evolution is driven by mass loss via the secondary's stellar wind, causing the orbit to further expand to the currently observed value. The fraction of this stellar wind attracted and accreted by the BH is too small to significantly influence the orbital evolution, but it is adequate to explain the X-Ray luminosity observed.
At the present time, after $\simeq$3.9$\,$Myr, M33 X-7 comprises a BH of  $\simeq$14.4$\,$M$_\odot$
 and an underluminous O-star of  $\simeq$64$\,$M$_\odot$ orbiting around their common center of mass in a slightly eccentric $\simeq$3.45$\,$day orbit. 

Our model is consistent with a natal nature of the BH's observationally inferred high spin\cite{LiuEtAl2008}. In fact, although it has been suggested that such a high spin is the result of a MT from the companion star to the BH through Roche-lobe overflow\cite{MorenoMendez2008}, given how more massive is the BH companion compared to the BH, such a phase could not have occurred: it would have been dynamically unstable and rapidly evolve into a merger of the binary components. Wind accretion is too weak to spin-up the BH to the current value if it were born spinning much slower. In our model, when the BH progenitor leaves the Main Sequence, the spins of the stars are expected to be synchronized with the orbit, and, assuming solid-body rotation on the Main Sequence, the inner parts of the primary's core carry enough angular momentum to explain the currently observed BH spin (the inner 15.5$\,$M$_\odot$ of the core carry $\simeq$5.2$\times$10$^{51}$$\,$gr$\,$cm$^2$$\,$s$^{-1}$ of angular momentum, while $\simeq$1.8$\times$10$^{51}$$\,$gr$\,$cm$^2$$\,$s$^{-1}$ is needed to explain the currently observed BH's spin). At the end of the Main Sequence, the inner core of the star is expected to rotationally decouple from the outer envelope and approximately retain the angular momentum of the central layers (ref. \cite{HirschiEtAl2005}, Table 4).

 We explore binary evolutionary sequences with different combinations of initial masses and orbital periods. We select as ``successful'' sequences those that eventually match all observed properties within 1$\sigma$ errors. All these sequences follow a path qualitatively very similar to the specific example described in detail here. The progenitors are constrained to host 96-99$\,$M$_\odot$ primaries, and 32$\,$M$_\odot$ (within 1$\,$M$_\odot$ uncertainty) secondaries in orbits with initial periods of 2.8-2.9$\,$days. The apparent puzzling underluminosity of the BH companion is due to two factors: (i) the orientation of the system with respect to our line of sight and associated projection effects reduce the star's measured luminosity (accounting for $\simeq$87\% of the underluminosity); (ii) the secondary  was not  born as a $\simeq$63-65$\,$M$_\odot$ star, but instead accreted much of its mass from the BH progenitor (accounting for $\simeq$13\% of the underluminosity) (see Supplementary Information). 

We note that the distance to M33 is more uncertain than 840$\pm$20 kpc, and some of the observed system properties vary if a different distance to the system is considered (see Table 1). Various studies in the literature have reported it in the range 750-1017$\,$kpc (see Supplementary Information). Considering this full range the progenitors are constrained to host primaries between 85-99$\,$M$_\odot$, secondaries between 28-32$\,$M$_\odot$, and initial orbital periods between 2.8-3.1$\,$days (see Figure 2). 

The different phases we described for M33 X-7's evolutionary past have been observed in a variety of other binary systems. For example, LMC  R136-38 hosts two non-interacting O-stars of $\simeq$57$\,$M$_\odot$ and $\simeq$23$\,$M$_\odot$ orbiting around each other every $\simeq$3.4$\,$days\cite{Massey2002}. LMC-SC1-105 comprises a $\simeq$31$\,$M$_\odot$ O-star that is transferring mass to its $\simeq$13$\,$M$_\odot$ companion\cite{Bonanos2009}. 
In the system WR46, a $\simeq$51$\,$M$_\odot$ Wolf-Rayet star orbits a $\simeq$60$\,$M$_\odot$ O-star companion
every $\simeq$6$\,$days. In particular, the various discoveries of Wolf-Rayet stars with 
O-star companions\cite{vanderHucht2001} validate our theoretical model, as they resemble
 the configuration of M33 X-7 less than 2$\,$Myr ago. Several of these binaries have been observed with orbital periods of a few days, and massive components (up to $\simeq$80$\,$M$_\odot$)\cite{Rauwetal2004}$^,$\cite{Bonanosetal2004}.

\begin{addendum}
 \item[Supplementary Information]is linked to the online version of the paper at www.nature.com/nature.
 \item We thank N. Ivanova, A. Heger, A. Cantrell, and C. Bailyn for useful discussions during the development of this project. 
 \item[Contributions] V.K. and B.W. conceived the study. F.V. led the project, performed the single and binary star evolution calculations, and developed the code to perform the orbital evolution after black hole formation. V.K. and B.W. collaborated with F.V. in each step of the project. E.G. maintained, updated and extended the stellar evolution code used, and collaborated with F.V. in performing the calculations. W.M.F. determined the correction to the star's luminosity and surface temperature due to the inclination of the system. T.F. led the theoretical analysis of the black hole spin. J.A.O. performed the analysis of M33 X-7's observational data with the ELC code for the full distance uncertainty. J.L. recalculated the black hole's spin at different distances. All authors discussed the results and made substantial contributions to the manuscript.
 \item[Competing Interests] The authors declare that they have no
competing financial interests.
 \item[Correspondence] Correspondence and requests for materials
should be addressed to F.V.
 \newline
~(email: francesca@u.northwestern.edu).
\end{addendum}

\begin{center}
\includegraphics[width = 1.0\textwidth]{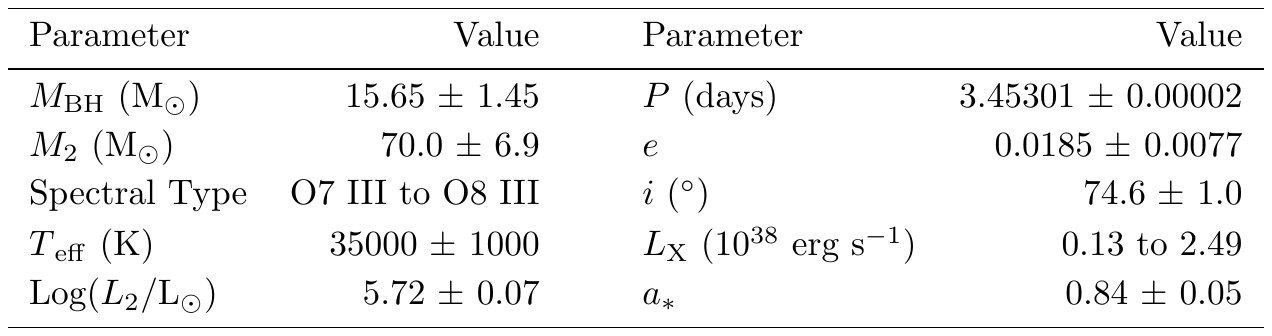}
\end{center}

\sffamily{
\begin{table}
\label{observedSystemParameters}
\caption{\sffamily Observed parameters for M33 X-7 for a reference distance of 840$\pm$20 kpc as adopted by the discovery team\cite{2007Nature}. The BH mass ($M_{\rm{BH}}$), its companion mass ($M_{\rm{2}}$), Spectral Type, effective temperature ($T_{\rm{eff}}$), and luminosity ($L_{\rm{2}}$), the orbital eccentricity ($e$), and inclination ($i$) are listed as reported by ref. \cite{2007Nature}. The orbital period ($P$) has been measured by ref. \cite{Pietsch2006}. The dimensionless spin parameter of the BH ($a_{\rm{*}}$) has been determined by ref. \cite{LiuEtAl2008} based on the X-ray continuum fitting method\cite{ZhangEtAl1997, LiEtAl2005, McClintockEtAl2006}.
The X-ray luminosity ($L_{\rm{X}}$) is derived from observations reported in refs. \cite{ParmarEtAl2001}$^,$\cite{PietschEtAl2004}$^,$\cite{Pietsch2006}$^,$\cite{ShporerEtAl2007}$^,$\cite{2007Nature}$^,$\cite{LiuEtAl2008}.
To account for variations in the X-Ray flux over different observations, we consider the lowest and highest  reported values, after we rescale each $L_{\rm{X}}$ to a M33 distance of 840$\pm$20$\,$kpc. If the full distance range of 750-1017$\,$kpc is adopted, using the ELC code of
Orosz and Hauschildt\cite{Orosz2000} we calculate that the masses are between 55-103$\,$M$_\odot$ and 13.5-20$\,$M$_\odot$ for the star and the BH, respectively, and the inclination is between 77$\degree$-71$\degree$. The logarithmic luminosity in solar units is then between 5.62-5.89\cite{2007Nature}. For each distance, the mass of the star can be derived from $M\rm_{2}$= -75.94 + 0.17 $\cdot$ $d$, and the inclination from  $i$ = 93.69 - 0.02 $\cdot$ $d$, where $d$ is in kpc, $M\rm_{2}$ in M$_\odot$ and $i$ in degrees. For each $M\rm_{2}$, the corresponding BH's mass in solar units can be calculated from $M\rm_{BH}$ = 6.19 + 0.13 $\cdot M\rm_{2}$, and the BH's spin from $a\rm_{*}$ = 0.31 + 0.049 $\cdot M\rm_{BH}$ - 0.001$\cdot M\rm_{BH}^2$. The rescaled X-Ray luminosity ranges from $\simeq$1$\times$10$^{37}$ to $\simeq$3$\times$10$^{38}$$\,$erg sec$^{-1}$.}
\label{observedSystemParameters}
\end{table}
}

\begin{center}
\includegraphics[width = 1.0\textwidth]{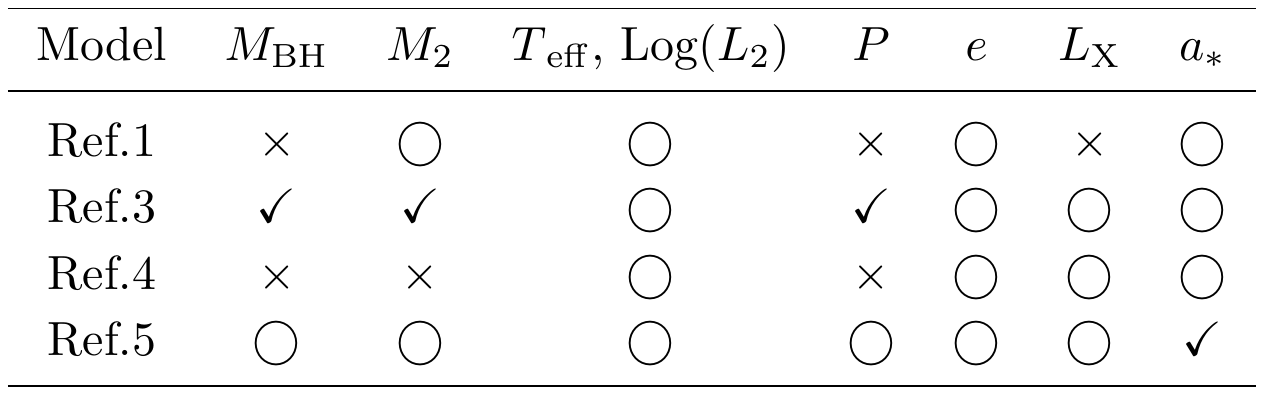}
\end{center}

\begin{table}
\label{previousModels}
\caption{\sffamily Models suggested for M33 X-7's formation. The various parameters are described in Table 1. The symbol "$\checkmark$"  means a parameter has been addressed and explained, "$\times$" a parameter has been addressed but not explained, and "$\bigcirc$" a parameter has not been addressed at all. To explain the tight orbit, ref. \cite{2007Nature} suggests that the progenitors underwent a common envelope (CE) phase during which the primary expanded to the point of engulfing its companion. Such a phase is known to lead to tight systems because energy can be transferred from the orbit to the CE, leading to a reduction of the binary separation, and ejection of the envelope\cite{TaamSandquist2000}. To form the observed BH, ref. \cite{2007Nature} requires that the CE begins after He core burning in the primary is completed. However, a CE in M33 X-7's case would likely evolve into a merger, because massive-star envelopes are tightly bound\cite{PodsiadlowskiEtAl2003}. Furthermore, for this model to succeed an unrealistically low stellar wind would be required. Ref.\cite{Abubekerov2009} suggests a phase of conservative MT from the BH progenitor to the companion that sets in at the end of the primary's Main Sequence, but this model only explains the observed masses and orbital period, failing to address the remaining observations. Ref.\cite{DeMinkEtAl2009} proposed rotationally induced mixing as a way to keep massive stars from expanding significantly during the Main Sequence, preventing MT or a merger until the primary becomes a Wolf-Rayet star. However, this evolutionary channel increases the star's luminosity above that of standard models, in contrast to the observed underluminosity of the star in M33 X-7. Ref.\cite{MorenoMendez2008} explains the observed BH's spin via a past Roche-lobe overflow phase from the star to the BH. However, given the extreme mass ratio between the components, such a phase would have evolved into a merger.}
\end{table}

\begin{center}
\includegraphics[width = 1.0\textwidth]{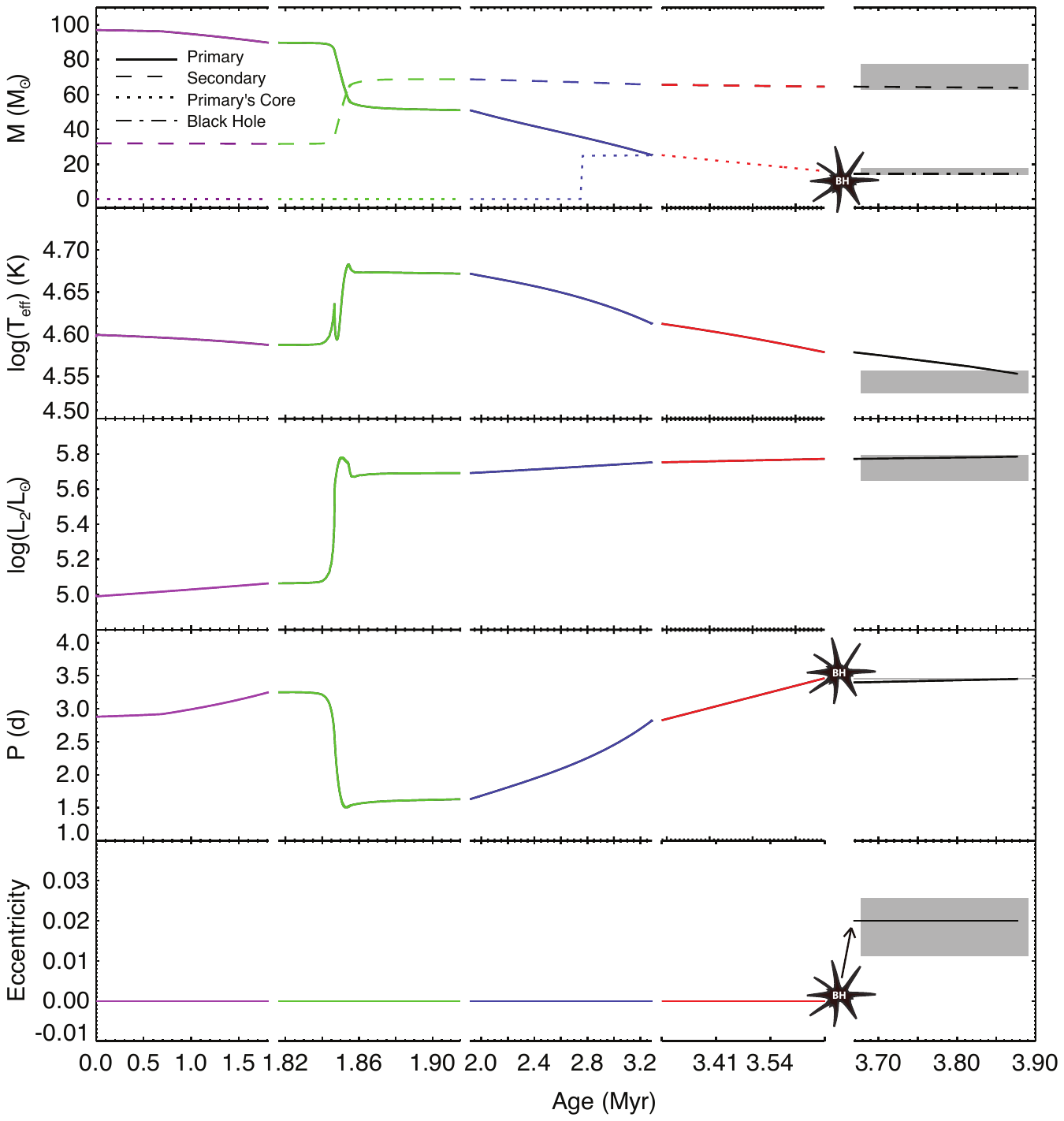}
\end{center}

\begin{figure}
\label{overallEvolution}
\caption{\sffamily Evolution of the orbital and stellar parameters of M33 X-7. From the top: masses ($M$), secondary's surface temperature [$\log(T_{\rm{eff}}$)], secondary's luminosity [$\log(L_{\rm{2}}$)], orbital period ($P$), and eccentricity ($e$). The different evolutionary stages are highlighted with different colors:  the beginning of the Main Sequence (purple), the MT phase (green), the end of the Main Sequence (blue), the core He burning phase for the BH progenitor (red), and the post BH formation phase (black) until the present time (note the non uniform x-axis).
The grey-shaded areas represent the observational constraints as reported in Table~\ref{observedSystemParameters}.
The sequence comprises a $\simeq$97$\,$M$_\odot$ primary ($M\rm_{1}$) and a $\simeq$32$\,$M$_\odot$ secondary ($M\rm_{2}$) 
in a $\simeq$2.9$\,$day orbit. At the onset of the MT phase (purple/green), $M\rm_{1}\simeq$89.7$\,$M$_\odot$, $M\rm_{2}\simeq$31.7$\,$M$_\odot$, and $P\simeq$3.25$\,$days. During the MT phase (green), the primary transfers conservatively $\simeq$37$\,$M$_\odot$ to the companion (see Supplementary Information for details about the MT). When the system detaches (green/blue) $M\rm_{1}\simeq$51$\,$M$_\odot$, $M\rm_{2}\simeq$68.7$\,$M$_\odot$, and $P\simeq$1.6$\,$days. At the end of the primary's Main Sequence (blue/red) $M\rm_{1}\simeq$25.2$\,$M$_\odot$, $M\rm_{2}\simeq$65.6$\,$M$_\odot$, and $P\simeq$2.8$\,$days. Considering evolutionary models of He stars (see Supplementary Information), we find that a He star of $\simeq$25$\,$M$_\odot$ burns He at its center for $\simeq$0.38$\,$Myr and loses $\simeq$9.1$\,$M$_\odot$ in its Wolf-Rayet wind. During this time the secondary loses $\simeq$1.1$\,$M$_\odot$ (red). 
Before BH-formation (red/black) $M\rm_{1}\simeq$16$\,$M$_\odot$,  $M\rm_{2}\simeq$64.5$\,$M$_\odot$, and $P\simeq$3.46$\,$days. For the case shown, at the primary's collapse a kick of 120 km/s is imparted to the newly born BH (see Supplementary Information for the allowed kicks).}
\end{figure}

\begin{center}
\includegraphics[width = 1.0\textwidth]{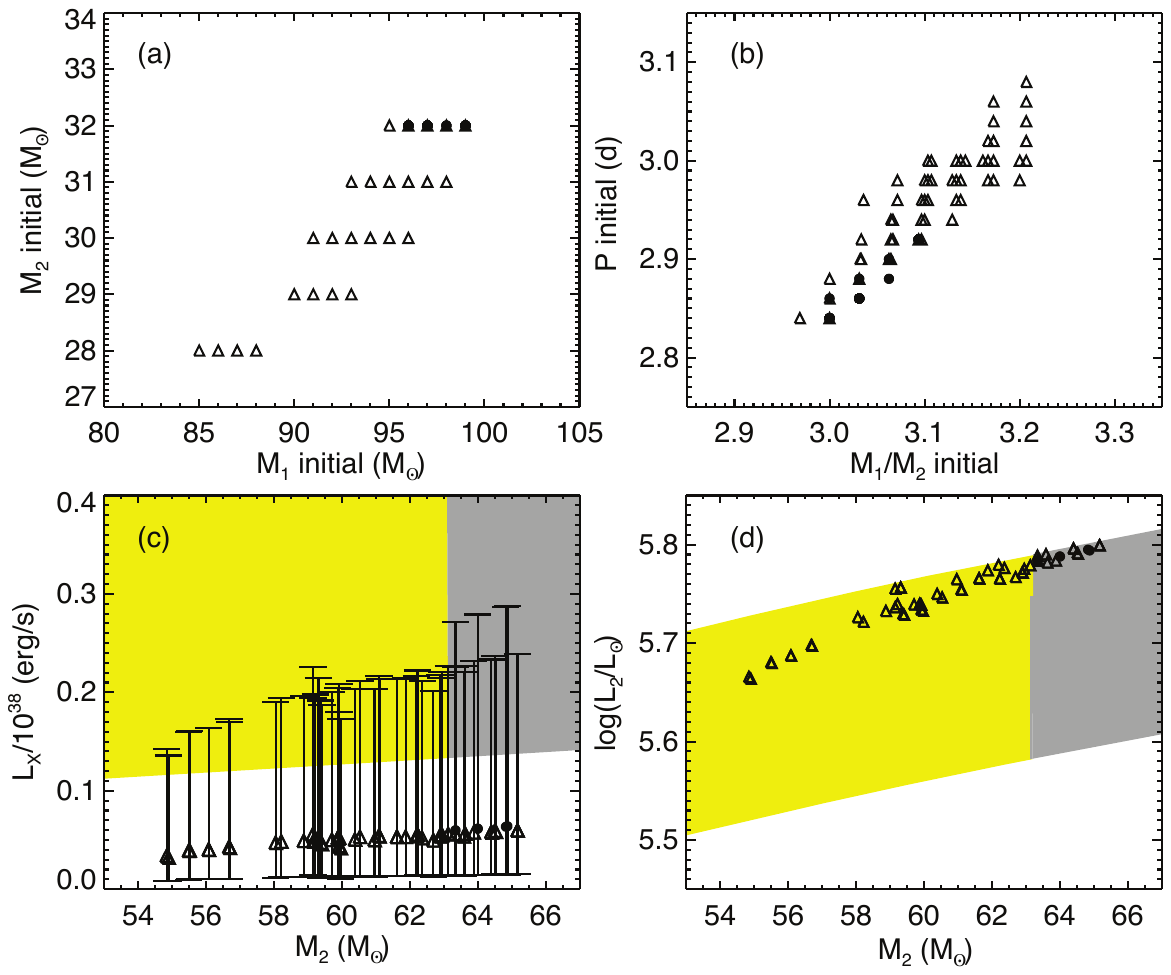}
\end{center}

\newpage
\begin{figure}
\label{ParametersAtPresent}
\caption{\sffamily Progenitor properties and current luminosity. The circles and triangles are the results of detailed binary star evolution calculations for all successful sequences for a M33 distance of 840 $\pm$ 20$\,$kpc, and of 750-1017$\,$kpc, respectively. (a) possible masses for the progenitors; (b) possible initial orbital periods as a function of the mass ratio between the primary and the secondary; (c) BH X-Ray luminosity and (d) secondary's luminosity as a function of the secondary's mass at present. The grey and yellow shaded areas represent the observational constraints for a distance of 840 $\pm$ 20$\,$kpc, and of 750-1017$\,$kpc, respectively. 
$L_{\rm{X}}$ is calculated according to the Bondi and Hoyle accretion model\cite{BondiHoyle1944}. The error bars are derived from the uncertainties in the stellar wind parameters (see Supplementary Information), and depict the highest and lowest $L_{\rm{X}}$ values; they do not represent statistical 1$\sigma$ errors. The observational constraints on $L\rm_{2}$ are calculated given the dependence of $M\rm_{2}$ and $L\rm_{2}$ on the distance as described in the caption of Table~\ref{observedSystemParameters}, and accounting for uncertainties in the star's effective temperature, reddening, and apparent magnitude calculated through the ELC code.  According to our model, secondaries at present more massive than $\simeq$65$\,$M$_\odot$ fail to explain the observed luminosity. Some of the data points are omitted for clarity.}
\end{figure}

\end{document}


\maketitle
\begin{affiliations}
 \item Center for Interdisciplinary Exploration and Research in Astrophysics (CIERA) and Department of Physics and Astronomy, Northwestern University, 2145 Sheridan Road, Evanston, IL 60208, USA
 \item Department of Physics and Astronomy, McMaster University,  1280 Main Street West, Hamilton, Ontario, Canada L8S 4M1
 \item Department of Astronomy, San Diego State University, 5500 Campanile Drive, San Diego, CA 92182-1221, USA
 \item National Astronomical Observatories, Chinese Academy of Sciences, Beijing 100012, China
 \item Harvard-Smithsonian Center for Astrophysics, 60 Garden Street, Cambridge, MA 02138, USA
\end{affiliations}

\thispagestyle{empty}
\pagestyle{empty}

\nocite{2007Nature}
 \nocite{Pietsch2006}
 \nocite{Abubekerov2009}
 \nocite{DeMinkEtAl2009}
 \nocite{MorenoMendez2008}
 \nocite{UetAl2009}  
 \nocite{Verbunt1993}
 \nocite{TaurisHeuvel2003}
 \nocite{LiuEtAl2008}
 \nocite{HirschiEtAl2005}
 \nocite{Massey2002}
 \nocite{Bonanos2009}
 \nocite{vanderHucht2001}
 \nocite{Rauwetal2004}
 \nocite{Bonanosetal2004}
 \nocite{ZhangEtAl1997}
 \nocite{LiEtAl2005}
 \nocite{McClintockEtAl2006}
 \nocite{ParmarEtAl2001}
 \nocite{PietschEtAl2004}
 \nocite{ShporerEtAl2007}
 \nocite{Orosz2000}
  \nocite{TaamSandquist2000}
 \nocite{PodsiadlowskiEtAl2003}
  \nocite{BondiHoyle1944}
\newpage

\noindent This Supplementary Information provides details about the uncertainty of the distance to M33, 
the single and binary star evolution models, the binary evolution on the primary's Main Sequence throughout the mass transfer (MT) phase, the orbital evolution after
black hole (BH) formation, the He stars models, the correction to
the luminosity and temperature of the stellar component due to tidal and rotational
distortions and the inclination of the system, the correction to the luminosity of the stellar component due to the
partial-rejuvenation of the star after MT, the parameter space
considered, and the stellar wind model used.  It also contains six
related figures, and additional references.

\newpage

{\LARGE{\section{Supplementary Information}}}

\noindent\textbf{Uncertainty in the distance to M33:} given the considerable uncertainty in the distance to M33 found in the literature, we consider a compilation of 15 recent distance measurements obtained via different techniques (see Supplementary Figure~\ref{distanceM33}). 
We consider the full range of uncertainty by combining the highest and smallest values, and obtain that the distance lies in the range 750-1017$\,$kpc.

\noindent\textbf{Single and binary star evolution models:} the stellar evolution models were calculated with an up-to-date version of Eggleton's
stellar evolution code, STARS\cite{Eggleton1971}$^,$\cite{Eggleton1973}$^,$\cite{Polsetal1995}$^,$\cite{Eggleton2002}.
The STARS code solves the equations of stellar structure and the
reaction-diffusion equations governing the nuclear energy generation
rate simultaneously on an adaptive non-Lagrangian non-Eulerian grid.
We used the code in so-called TWIN mode, where both components of a binary are
evolved simultaneously. This is important if both stars evolve on comparable timescale, as is the case for the M33 X-7 progenitors.

For the thermonuclear reaction rates we use the
recommended values from NACRE\cite{NACRE}, with the exception of the
$^{14}\mathrm{N} (\mathrm{p}, \gamma)^{15}\mathrm{O}$ reaction, for which
we use the recommended rate from Herwig et al.\cite{Herwig2006} and 
Formicola et al.\cite{FormicolaEtAl2004}.
The adopted opacity tables are those of Pols et al.\cite{Polsetal1995}, which combine the
OPAL opacities from Rogers \& Iglesias\cite{RogersIglesias1992}, and the low temperature
molecular opacities from Alexander \& Ferguson\cite{AlexanderFerguson1994}. 
Conductivities come from Itoh et al.\cite{Itohetal1983} and Hubbard \& Lampe\cite{HubbardLampe1969}

The assumed heavy-element composition is scaled to the solar mixture of Anders \& 
Grevesse\cite{AndersGrevesse1989}. Chemical mixing due to
convection\cite{BohmVitense1958}$^,$\cite{Eggleton1973} and thermohaline 
mixing\cite{Kippenhahn1980}$^,$\cite{Stancliffe2007} is also taken into account. For this work
we furthermore added a prescription for semi-convection based on the work of
Langer et al.\cite{Langeretal1983} All models are computed with a mixing-length to
pressure scale height ratio $l/H_P=2.0$. 
Differential rotation and mixing due to meridional circulation are not
taken into account. Although mixing due to meridional circulation can be
very important, the mixing efficiencies are also very uncertain and have to
be tuned to observations. We did not wish to include extra free parameters
in our model, which is intended to be the simplest model we can make that
fits the observations.

During the Main Sequence phase of the primary and for the secondary we use the mass loss prescription of Vink et al.\cite{Vink2001}. When the surface H
abundance by mass fraction in the primary drops below $0.4$ we switch to the Wolf-Rayet
prescription of Nugis \& Lamers\cite{NugisLamers2002} with the metallicity scaling
determined by Vink \& de Koter \cite{VinkKoter2005}. If the effective temperature drops
below 10,000$\,$K the Vink et al.\cite{Vink2001} prescription loses its
validity and we fall back to the mass loss rate of de Jager et al.\cite{deJager1988}.

\noindent\textbf{Binary evolution on the primary's Main Sequence, throughout the MT phase:} the efficiency of the process of mass accretion during a phase of MT between the components of a binary system is still an open question, and in massive close binaries there is evidence for both quasi-conservative and highly non-conservative evolution\cite{LangerEtAl2003}. Calculations of massive binary evolution have been carried out by various authors with different assumptions for the mass accretion efficiency, and the clear outcome of these studies is that different mass accretion efficiencies might be needed to explain different observations.
For example, Petrovic et al.\cite{PetrovicEtAl2005} explored the evolutionary history of three of the about 20 Wolf-Rayet binary systems known in the catalogue of van der Hucht\cite{vanderHucht2001}, and they concluded that if the considered systems underwent a phase of stable MT, in order to match the observed systems properties a large amount of mass must have left the system. On the other hand, De Mink et al.\cite{deMinkEtAl2007} calculated detailed evolutionary tracks with primary masses between 3.5-35$\,$M$_\odot$, mass ratios between the primary and the secondary components from $\simeq$1 to $\simeq$2.2, and orbital periods of a few days assuming both conservative and non-conservative MT. A systematic comparison of these evolutionary models with a sample of 50 double-lined eclipsing binaries in the Small Magellanic Cloud revealed that, for the 17 systems well matched by the models, ``no single value for the efficiency of mass accretion can explain all systems'', and they found good agreement between the model and the observed systems properties for accretion efficiencies up to 1 (conservative evolution).
Given that we can find successful M33 X-7 formation sequences assuming quasi-conservative MT (with the only mass loss from the system being due to the stellar wind of both components) we conclude that there is no reason to invoke non-conservative MT, and introduce more model parameters.

A description of some of the relevant physical quantities involved in the MT phase for the binary sequence described in detail in the Letter are presented in Supplementary Figure \ref{DetailedEvolution}.

\noindent\textbf{Orbital evolution after BH formation:} we study the orbital evolution of M33 X-7 from the moment at which the BH is formed to the present time, by examining how the orbital separation, eccentricity, and spin frequency of the stellar component change in time. The orbital and spin evolution calculation accounts for the following physical mechanisms: tidal torques between the binary components, mass loss from the system via stellar wind, changes in the stellar radius during the Main Sequence lifetime of the companion star, binary angular momentum loss due to gravitational radiation, and accretion from the companion's stellar wind onto the BH. The BH is considered as a point mass and the relevant ordinary differential equations are integrated.

The tidal torques act to synchronize the rotational motion of each star with the orbital motion, and to circularize the orbit. The tidal evolution is calculated in the standard weak-friction approximation\cite{Zahn1977}$^,$\cite{Zahn1989}, following the formalism of Zahn\cite{Zahn1975}$^,$\cite{Zahn1977} and Hut\cite{Hut1981}, for stars with a radiative envelope. We assume that radiative damping is the only source of dissipation. Specifically, we integrate numerically the set of differential equations as presented in Belczynski et al.\cite{Belczynski2008}, with the only modification being in the second-order tidal coefficient $E_2$. 
For this coefficient, we adopt stellar models from Claret\cite{Claret2006} for masses of $\sim63\,$M$_\odot$ and $\sim79.5\,$M$_\odot$, and derive $E_2 = - 5.4566 - 7.37243\cdot t\rm_{MS}^{4.50562}$, where $t\rm_{MS}$ is time in units of the main sequence lifetime. The fitting formula has only a very weak dependence on the initial mass for the considered range.

Wind mass loss leads to an increase of the orbital separation and, together with the expansion of the star on the Main Sequence, affects the stellar spin.
The evolution of the orbital separation driven by stellar wind is calculated following Eggleton\cite{EggletonBook2006}, while the change in the rotational frequency is derived assuming conservation of spin angular momentum of the star. The wind mass loss tends to spin up the star, while the increase in the stellar radius has the opposite effect.

Emission of gravitational radiation acts to circularize the orbit and, together with accretion from the stellar wind onto the BH, shrinks the orbit. The evolution of the orbital separation and eccentricity due to gravitational radiation is calculated following Junker \& Schaefer\cite{Junker1992}. The accretion efficiency is calculated according to Bondi \& Hoyle\cite{BondiHoyle1944} (for details about the parameters used for describing the stellar wind see below). For the specific case of M33 X-7, both these mechanisms do not significantly influence the orbital evolution. 
In particular, the time scale for gravitational radiation is longer than the timescales relative to the other physical effects mentioned above, and the calculated accretion efficiency is extremely small ($\sim10^{-4}$). 

For each time step during the orbital evolution we calculate the Roche-lobe radius of the star at periastron\cite{SepinskyEtAlNoFred2007}. Considering that a phase of MT via Roche-lobe overflow during this evolutionary stage would have been dynamically unstable, and motivated by uncertainties in the definition of the stellar radius inherent to different models\cite{Valsecchi2009},  the maximum value of the radius is set equal to the Roche-lobe radius. 

\noindent\textbf{He star models:} a common problem among most stellar evolution codes occurs when trying to fully remove the H envelope of a massive star that has a H burning shell right outside the He burning core. As the code tries to remove the last bit of the H envelope ($<$1$\%$ of the total stellar mass) it reaches inside the H burning shell, causing numerical instabilities. One way to deal with this problem is to dramatically increase the spatial resolution of the simulations (by a factor $>$10). However, this results in unrealistically long computational times. An alternative approach, which we adopt here, is to stop calculations when this phase occurs and restart from a He star model. This avoids the short, numerically challenging phase without losing any information. 

Using the code STARS\cite{Eggleton1971}$^,$\cite{Eggleton1973}$^,$\cite{Polsetal1995}$^,$\cite{Eggleton2002} we create models of He stars with the same input physics used for the single and binary star evolution calculations, and with masses ranging from 3$\,$M$_\odot$ to 25$\,$M$_\odot$. We then evolve each model until the exhaustion of He in the core to determine the duration of the core He burning phase ($t\rm_{He}$) and the corresponding amount of mass lost ($\Delta M$) as a function of the initial mass of the He star ($M_{\rm{He, i}}$). The results are shown in Supplementary Figure \ref{HeModels}. From these models we derive the following relations:

\begin{equation}
\label{eq:He-burn-time}
t\rm_{He} =0.323221 + 6.24256\cdot M\rm_{He, i} ^{-1.45762}
\end{equation}

\begin{equation}
\label{eq:He-Mass-Loss}
\Delta M = 
\begin{cases} 0.282 - 0.28\cdot M\rm_{He, i} + 0.052 \cdot M\rm_{He, i}^2 - 0.00102\cdot M\rm_{He, i}^3    &    M\rm_{He, i} \leq 17  M_\odot \\
- 23.5355 + 10.262 \cdot \ln(M\rm_{He, i})   &   M\rm_{He, i} > 17  M_\odot \\
\end{cases}
\end{equation}

where $t\rm_{He}$ is in Myr, and $M\rm_{He, i}$ and $\Delta M$ are in M$_\odot$.

\noindent\textbf{Correction to the luminosity and temperature due to
  tides, rotation and inclination:} the shape of the star in M33 X-7
is distorted by rotation and tides.  This distortion causes the
temperature to vary over the surface of the star; the equatorial
regions are colder than the poles (this is a consequence of the Von
Zeipel theorem, which relates the effective temperature to the
one-fourth power of the surface gravity).  We use the ELC code of
Orosz and Hauschildt\cite{Orosz2000} to fit a surface temperature map
to the light curve and radial velocity curve.  Using this surface
temperature map, we find that
\begin{equation}
\langle T_{\rm eff} \rangle = 0.954 T_{\rm polar}.
\end{equation}
Here $\langle T_{\rm eff} \rangle$ is the flux-averaged effective
temperature, taking into account the inclination of 74.5 degrees
reported by Orosz et al.\cite{2007Nature}.  $T_{\rm polar}$ is the polar temperature; the 
polar temperature is the maximum temperature over the surface because rotation and tides have
the smallest effect at the poles.

The effective temperature range of 34000 to 36000 K reported by Orosz
et al.\cite{2007Nature} is a measurement of $\langle T_{\rm eff}
\rangle$.  However, our stellar models do not incorporate the effects
of tides or rotation on the surface effective temperature, so a direct
comparison of the model surface effective temperature to the measured
$\langle T_{\rm eff} \rangle$ is inappropriate.  Because the tides and
rotation have the smallest effect on the polar regions of the star, we
choose to compare the modeled surface effective temperature
with $T_{\rm polar} = \langle T_{\rm eff} \rangle / 0.954$.

The tidal and rotational distortions of the star also have an effect
on the observed luminosity.  The true luminosity of the star is an
integral over the surface of the local flux density (given by the
Stefan-Boltzmann law):
\begin{equation}
  L_{\rm true} = \int_S \sigma T_{\rm eff}^4 dA.
\end{equation}
The luminosity quoted by Orosz et al.\cite{2007Nature}, $\log(L_{\rm
  obs}/L_\odot) = 5.72 \pm 0.07$, is based on the quoted visual
magnitude, $V = 18.9 \pm 0.05$, of the star at an inclination of $i = 74.6$ degrees. 
Given a surface temperature map, we can compute the 
luminosity that would be inferred from the orbit-averaged emission at an inclination of 74.6 degrees:
\begin{equation}
\label{eq:Average-L}
 L_{\rm avg} = \frac{A}{A_{\rm vis}} \int_{S_{\rm vis}} \sigma T_{\rm
   eff}^4 \cos\left( \theta - i \right) dA.
\end{equation}
Here $S_{\rm vis}$ is the subset of the stellar surface visible at an
inclination of 74.6 degrees, $A_{\rm vis}$ is the corresponding area,
$A$ is the total surface area of the star, and $\theta$ is the angle
of the surface normal with respect to the vertical axis.  The
$\cos(\theta - i)$ factor accounts for the relative orientation of the
surface normal to the line of sight.  Referring to Figure 3(b) of
Orosz et al.\cite{2007Nature}, the average visual magnitude of the
light curve is $V = 18.87$; the $0.03$ magnitude difference between
the lightcurve average and the quoted visual magnitude corresponds to
$\Delta \log(L) = 0.01$.  So, we have
\begin{equation}
  \label{eq:true-Luminosity}
  \log (L_{\rm avg}/L_\odot) = \log (L_{\rm obs}/L_\odot) + 0.01.
\end{equation}
Based on the ELC surface temperature map and equation
\eqref{eq:Average-L}, we calculate
\begin{equation}
  \log(L_{\rm true}/L_\odot) = \log (L_{\rm avg}/L_\odot) + 0.13,
 \end{equation} 
and therefore 
\begin{equation}
  \label{eq:true-Luminosity}
  \log(L_{\rm true}/L_\odot) = \log (L_{\rm obs}/L_\odot) + 0.14.
\end{equation} 
This luminosity correction is independent of the absolute surface
temperature provided the surface temperature profile remains fixed
throughout the observed temperature range $34000 \leq \langle T_{\rm
  eff} \rangle \leq 36000$ K.  The observed luminosity is lower than
the true luminosity because the inclination of the system implies that
we are looking preferentially at the colder equatorial regions of the
star.  When comparing model luminosities with observations we use
$L_{\rm true}$ from equation \eqref{eq:true-Luminosity}, not $L_{\rm
  obs}$ from Orosz et al.\cite{2007Nature}; this amounts to a decrease in the
  model luminosities of 0.14.

\noindent\textbf{Correction to the luminosity due to partial-rejuvenation of the secondary:} since the work of Hellings\cite{Hellings1983,Hellings1984} on the evolution of Main Sequence mass-accreting secondaries, it is generally assumed that the accretion of matter via Roche-lobe overflow leads to so-called "rejuvenation" of the star. The central H abundance of the accreting secondary increases, and its internal chemical structure becomes almost identical to the structure of a $single$ star of the corresponding mass. Ten years after Hellings, Braun and Langer\cite{BraunLanger1995} showed that rejuvenation does not always occur and that the result of mass accretion might be a star with a chemical structure unlike that of an originally single star. One of the most influential parameters that controls this effect is the semiconvective mixing efficiency that, in turn, depends on the criterion for convection used in the stellar model. Hellings adopted the Schwarzschild criterion, according to which the semiconvective mixing efficiency is infinite, while Braun and Langer used the Ledoux criterion with a finite value for this parameter. Braun and Langer showed that, despite the increase in luminosity as a result of mass accretion, the non-rejuvenated models appear to be underluminous for their new mass during the remaining Main Sequence evolution.
Following Braun and Langer, we use the Ledoux criterion for convection in our detailed single and binary evolution calculations, using primordial composition for the transferred material, and we adopt for the semiconvective efficiency parameter a value of $\alpha_{\rm{sc}} = 0.0025$ (which was calibrated using some of the results reported by Braun and Langer\cite{BraunLanger1995}). Our results confirm that the rejuvenation of the secondary component after MT was at most partial (see Supplementary Figure \ref{nonRejuvenation} for an example).
  
\noindent\textbf{The parameter space:} keeping in mind that the short observed orbital period cannot be the result of a common envelope phase from the BH progenitor to the companion star, we explore the evolution of binary systems that start their life already in a tight orbit, hence undergoing a MT phase during the core H burning phase of the primary component. We perform the binary evolution calculations from the Zero-Age Main Sequence until the end of the primary's Main Sequence considering various combinations of initial masses and  orbital periods. Specifically, we evolve primaries and secondaries with masses between 20-130$\,$M$_\odot$ and 10-100$\,$M$_\odot$, respectively, and initial orbital periods ranging from 1 to 10$\,$days. Based on the observed masses of the two components we use a different density of models in different regions of the parameter space.  

Since we consider MT during the core H burning phase of the primary, we perform a first scan of the data rejecting the sequences where the primary overfills its Roche-lobe only after the end of its Main Sequence. Furthermore, given the high mass of the BH companion, we also exclude the sequences where the secondary ends up transferring mass to the primary after having accreted mass from it. After the end of the primary's Main Sequence, given that the star is a He star we use equations \eqref{eq:He-burn-time} and \eqref{eq:He-Mass-Loss} to calculate the amount of mass lost from the binary components, and the corresponding change in orbital period during the core He burning phase of the primary until collapse. We reject the sequences where the masses of the two stars at the end of the core He burning phase of the primary are lower than the minimum observed values for the minimum distance to M33 of 750$\,$kpc. Since no episodes of Roche-lobe overflow can have occurred from the collapse of the primary until the present time, we then evolve each secondary as a single star, and exclude the sequences where the mass of the star does not fall within the observed range when the model matches the observed effective temperature and luminosity. After the primary's collapse, we consider a variety of orbital configurations by scanning the parameter space made up of the kick magnitude ($V_{\rm{k}}$), orbital separation ($a_{\rm{postBH}}$), and eccentricity ($e_{\rm{postBH}}$).
Specifically, we consider isotropic kicks between 0-1300 km/s, orbital separations between 0-100$\,$R$_\odot$, and eccentricities between 0 and 1. The requirement that the system must remain bound after the BH is formed, and that the direction of the kick must be real impose constraints on the pre- and post-  BH formation orbital parameters\cite{Kalogera1996,WillemsEtAl2005}. We then study the orbital evolution after the formation of the BH, and interrupt the calculation when the orbital period crosses the observed value; then the eccentricity of the orbit must fall within the observed 1$\sigma$ range.

Finally, of the sequences that fulfill all the above requirements, we compare the stellar radius at present with the distance from the center of the star to the point through which mass would flow from the star to the BH in case of Roche-lobe overflow\cite{SepinskyetalWithFred2007}. 
We reject the sequences where the secondary companion is transferring mass to the BH at the present time. Supplementary Figure~\ref{MassesNow} shows the masses of the components at present according to our model. Supplementary Figure~\ref{SNparams} shows the allowed BH kicks, BH progenitor masses, orbital separations and eccentricities post-BH formation, for all the successful sequences. 

According to our model,  the BH progenitor mass lies within the observed range for the BH mass, and, hence, no baryonic mass is ejected at collapse. Furthermore, the allowed eccentricities post-BH formation are constrained to be between 0.012-0.026. On one hand, the lack of mass ejection at BH formation, and the small induced eccentricity, could imply that the BH did not received a high kick at formation. On the other hand, due to the lack of kinematic information, we can not exclude kicks as high as $\sim$850$\,$km/s. This apparent discrepancy is explained by the fact that the kick is constrained to point mostly orthogonal to the orbital plane. In this case, a higher kick results in a more tilted orbit, while the orbital eccentricity does not change significantly, but enough to explain the observed eccentricity of 0.0185$\pm$0.0077. An upper limit to the magnitude of the kick is given by the observationally inferred positive spin of the BH. Kicks higher than $\sim$850$\,$km/s would flip the orbital plane and that would result in a negative value for the BH's spin. 

\noindent\textbf{Stellar wind model and X-Ray luminosity:} to determine the stellar wind properties that enter the Bondi \& Hoyle\cite{BondiHoyle1944} accretion model we follow Lamers \& Cassinelli\cite{LamersCassinelli1999}, and we adopt a velocity law of the type 
\begin{equation}
v(r) = v_0 + (v_{\infty} - v_0)\left( 1- \frac{R}{r}\right)^{\beta},
\end{equation}
where $v_0$ is the escape velocity at the stellar surface, $v_\infty$ is the velocity of the wind at infinity, $R$ is the photospheric radius, and $\beta$ is an index which typically ranges from 0.8 to 1.2 (we use $\beta = 1.0\pm 0.2$).
The escape velocity is defined as
\begin{equation}
v_0 = \left[\frac{2GM(1-\Gamma)}{R}\right]^{1/2},
\end{equation}
where
\begin{equation}
\Gamma = 7.66 \times 10^{-5}\sigma_{\rm{e}} \left(\frac{L}{L_{\odot}}\right)\left(\frac{M_{\odot}}{M}\right).
\end{equation}
For the electron scattering coefficient $\sigma_e$ we use
\begin{equation}
\sigma_{\rm{e}} = 0.401\frac{1+q\epsilon}{1+3\epsilon}, 
\end{equation}
where q is the fraction of He$^{++}$, and $\epsilon = He/(H+He)$. Following Lamers \& Leitherer \cite{LamersLeitherer1993}, we adopt  $\epsilon = 0.15 \pm 0.05$, which is appropriate for an O-type stars of spectral class III, $q$ = 1 if the effective temperature of the star ($T_{\rm{eff}}$) is $\geq 35,000 \,$K, or $q$ = 1/2 if $30,000\leq T_{\rm{eff}}<35,000$. For the velocity of the wind at infinity we adopt $v_\infty/v_0 = 3.085\pm 1.075$, which is again appropriate for a star of the spectral class observed for the companion star in M33 X-7\cite{LamersSnowLin1995}.
To calculate the mass accretion rate via stellar wind according to the Bondi \& Hoyle accretion model, we follow Belczynski et al.\cite{Belczynski2008} (and references therein):
\begin{equation}
\dot{M}_{\rm{acc, wind}} = - \frac{F_{\rm{wind}}}{\sqrt{1 - e^2}}\left(\frac{GM_{\rm{acc}}}{V^2_{\rm{wind}}}\right)^2\frac{\alpha_{\rm{wind}}}{2a^2}\frac{\dot{M}_{\rm{don, wind}}}{(1+ V^2)^{3/2}}.
\end{equation}
We use $\alpha_{\rm{wind}} = 3/2$, and $F_{\rm{wind}} = 1$. If $\dot{M}_{\rm{acc, wind}}$ exceeds $0.8 \dot{M}_{\rm{don, wind}}$, $F_{\rm{wind}}$ is set such that $\dot{M}_{\rm{acc, wind}}$ = $0.8 \dot{M}_{\rm{don, wind}}$.
$M_{\rm{acc}}$ is the mass of the accreting component,  $\dot{M}_{\rm{don, wind}}$ is the mass loss rate via wind, and $V^2 =V^2_{\rm{acc, orb}}/V^2_{\rm{wind}}$. The orbital velocity of the accretor is given by $V^2_{\rm{acc, orb}} = G(M_{\rm{acc}}+M_{\rm{don}})/a$, and $V^2_{\rm{wind}}$ is the squared of the velocity of the wind at $r = a$, where $a$ is the orbital separation.
Following Belczynski et al.\cite{Belczynski2008}, the bolometric luminosity is calculated from the mass accretion rate
\begin{equation}
L_{\rm{bol}} = \epsilon \frac{GM_{\rm{acc}}\dot{M}_{\rm{acc, wind}}}{R_{\rm{acc}}},
\end{equation}
where $\epsilon$ gives a conversion efficiency of gravitational 
binding energy to radiation associated with accretion onto a compact object, and is equal to 0.5 for accretion onto a BH. $R_{\rm{acc}}$ is the radius of the accretor, and we calculate it from the observationally inferred spin following Bardeen et al.\cite{BardeenEtAl1972}.
Finally, the X-Ray luminosity is calculated from the bolometric accretion luminosity via\cite{Belczynski2008}:
\begin{equation}
L_{\rm{X}} = \eta_{bol} L_{\rm{bol}};
\end{equation}
where we use $\eta=0.8\pm 0.1$.

\newpage
{\LARGE{\section{Acknowledgements}}}
\noindent This work was partially supported by NSF grants AST--0908930 and CAREER AST--0449558 to VK.  TF is supported by a Northwestern University Presidential Fellowship. Simulations were performed on the computing cluster {\tt Fugu} available to the Theoretical Astrophysics group at Northwestern and partially funded by NSF grant PHY--0619274 to VK. 

{\LARGE{\section{Supplementary Notes}}}
\bibliography{myBibtex}{}
\newpage
{\LARGE{\section{Supplementary Figures And Legends}}}
\begin{figure}
\begin{center}
\label{distanceM33}
\resizebox{12cm}{!}{\includegraphics{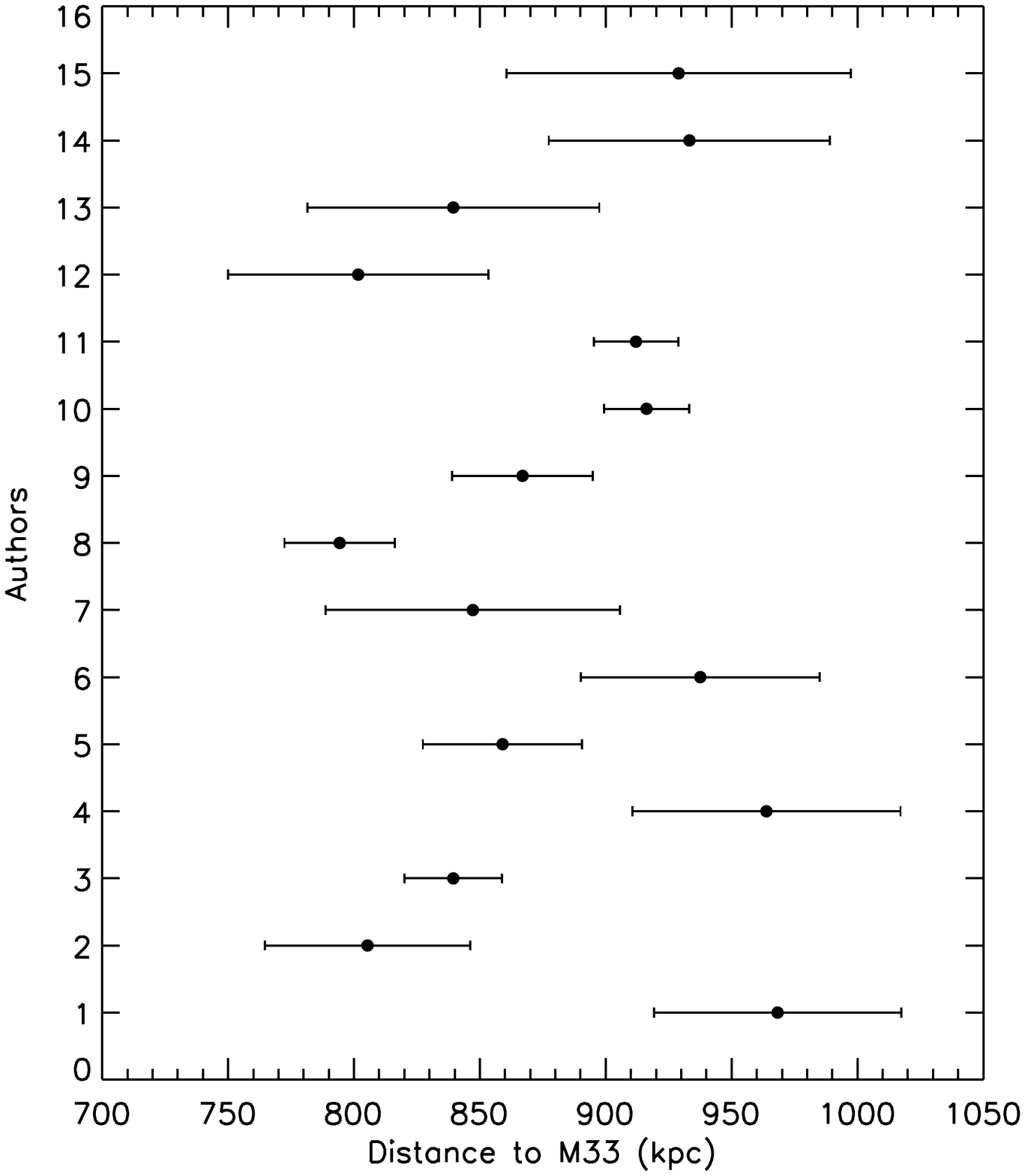}}
\end{center}
\caption{\textbf{ Recent measurements of the distance to M33.} From the bottom, 1: U et al.\cite{UetAl2009}, 2: Scowcroft et al.\cite{ScowcroftEtAl2009}, 3: Orosz et al.\cite{2007Nature}, 4: Bonanos et al.\cite{BonanosEtAl2006}, 5: Sarajedini et al.\cite{SarajediniEtAl2006}, 6: Ciardullo et al.\cite{CiardulloEtAl2004}, 7: Galleti et al.\cite{GalletiEtAl2004}, 8: McConnachie et al.\cite{McConnachieEtAl2004}, 9: Tiede et al.\cite{TiedeEtAl2004}, 10 and 11: Kim et al.\cite{KimEtAl2002}, 12: Lee et al.\cite{LeeEtAl2002}, 13: Freedman et al.\cite{FreedmanEtAl2001}, 14: Pierce et al.\cite{PierceEtAl2000}, 15: Sarajedini et al.\cite{SarajediniEtAl2000}. The last 12 estimates have been calculated from the distance moduli listed in the compilation given by Bonanos et al.\cite{BonanosEtAl2006}. We do not include the distance measured from Brunthaler et al.\cite{BrunthalerEtAl2005} because of its big uncertainty. In fact, both the statistical and systematic errors are quite large compared to other methods. A longer time baseline of observations would help reduce the statistical error (Bonanos 2010, private communication).} 
\label{distanceM33} 
\end{figure}

\begin{figure}
\begin{center}
\label{DetailedEv}
\resizebox{12cm}{!}{\includegraphics{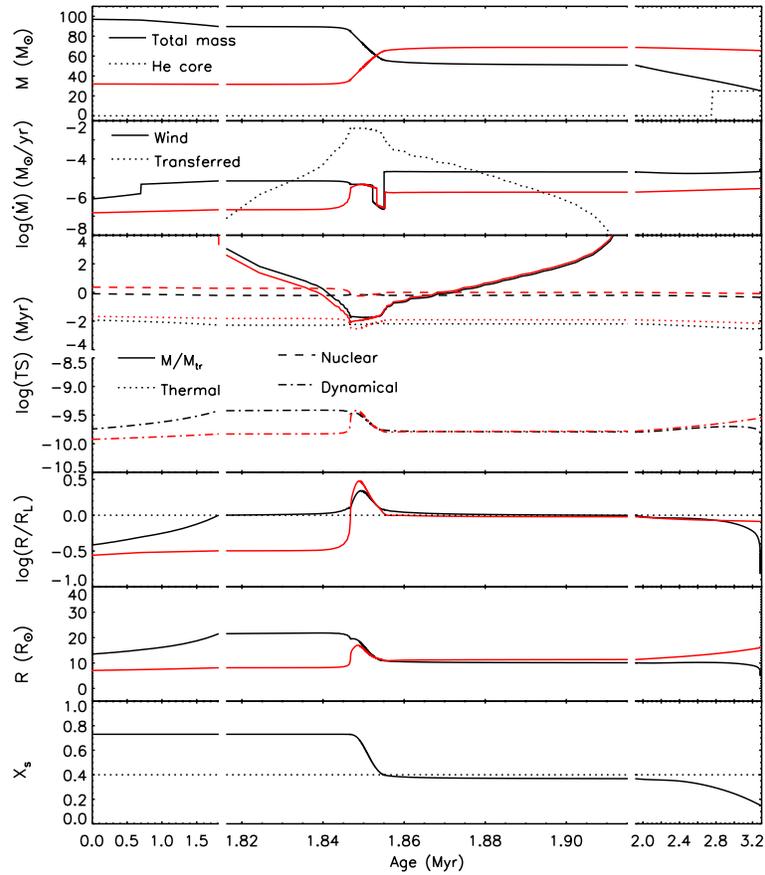}}
\end{center}
\caption{\textbf{The conservative MT phase.} From the top: masses of the components, mass rate of change ($\dot{M}$), typical stellar evolution timescales ($TS$), Roche-lobe overflow filling factor ($R/R_{\rm{L}}$), stellar radius ($R$), and primary's H surface abundance by mass fraction ($X_{\rm{s}}$) as a function of time for the evolutionary sequence described in the Letter. Black and red lines indicate the primary and secondary component, respectively. $\dot{M_{\rm{tr}}}$ is the mass transfer rate. The MT (or Roche-lobe overflow) phase is denoted by a positive value for  $\log{(R/R_{\rm{L}})} $, where $R_{\rm{L}}$ is the star's Roche-lobe radius (the dotted line in the corresponding plot is given as a reference for $\log{(R/R_{\rm{L}})} = 0$). When $X_{\rm{s}}$ drops below $0.4$ the Wolf-Rayet regime is entered (the dotted lines is given as a reference to this value). Given the short duration of the MT phase with respect to the total Main Sequence lifetime of the primary star, we use a non uniform x-axis.  The $\simeq$97$\,$M$_\odot$ primary evolves and expands faster than the secondary, and after $\sim$1.8$\,$Myr overfills its Roche-lobe and begins transferring part of its envelope to the companion. Within the first $\sim$37,000$\,$yr into the MT phase the orbital period decreases and, as a result, the rate at which the primary transfers mass increases from $\sim$10$^{-8}$$\,$ M$_{\odot}$/yr to $\sim$3$\cdot$ 10$^{-3}$$\,$ M$_{\odot}$/yr. At this time, the MT and mass accretion timescales ($M_{\rm{1}}/\dot{M_{\rm{tr}}}$ and $M_{\rm{2}}/\dot{M_{\rm{tr}}}$, respectively) become comparable to the stars' thermal timescales.  This brings the components out of thermal equilibrium (the components remain in hydrostatic equilibrium). The secondary overfills its Roche-lobe radius as well for a short time ($\sim$9,000$\,$yr) without transferring any mass, but when the component masses become equal and the orbit begins expanding, both 
stars recover their thermal equilibrium, and the secondary detaches. Once the secondary is detached and 
the orbital period (and Roche-lobe radius) are increasing, the primary keeps 
transferring mass for $\sim$59,000$\,$yr at a decreasing rate. When 
$X_{\rm{s}}$ of the primary drops below $0.4$ the star enters the 
Wolf-Rayet regime, and the corresponding stronger stellar wind interrupts the MT, and removes the remaining stellar envelope to expose the He core.}
\label{DetailedEvolution} 
\end{figure}

\begin{figure}
\begin{center}
\label{HeModels}
\resizebox{12cm}{!}{\includegraphics{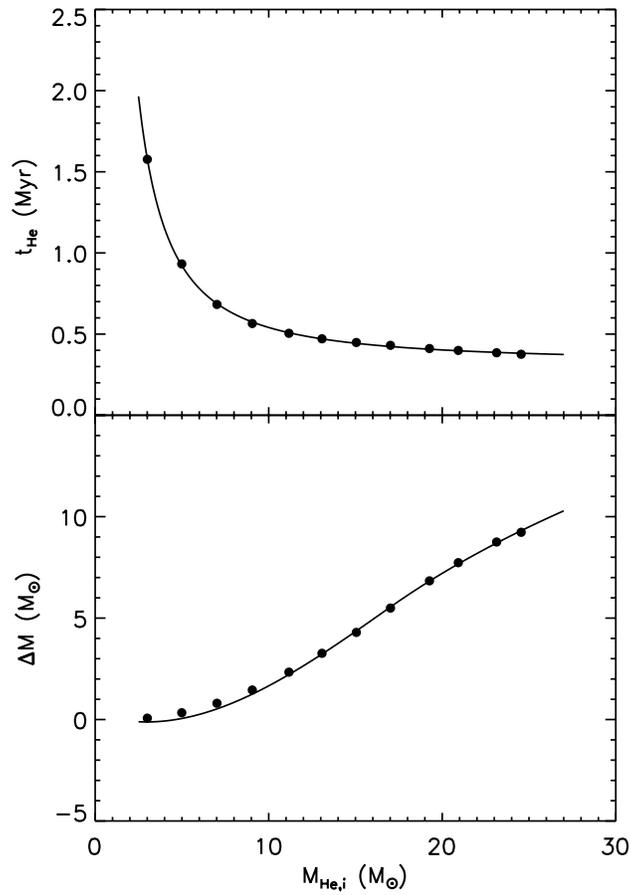}}
\end{center}
\caption{\textbf{Core He burning phase for He stars with different initial masses.} $Top$: duration of the core He burning phase as a function of the initial mass. $Bottom$: amount of mass lost during the core He burning phase as a function of the initial mass. The dots represent He star models, while the solid lines are the fits in equations \eqref{eq:He-burn-time} and \eqref{eq:He-Mass-Loss}.}
\label{HeModels} 
\end{figure}
\begin{figure}
\begin{center}
\label{nonRejuvenation}
\resizebox{10cm}{!}{\includegraphics{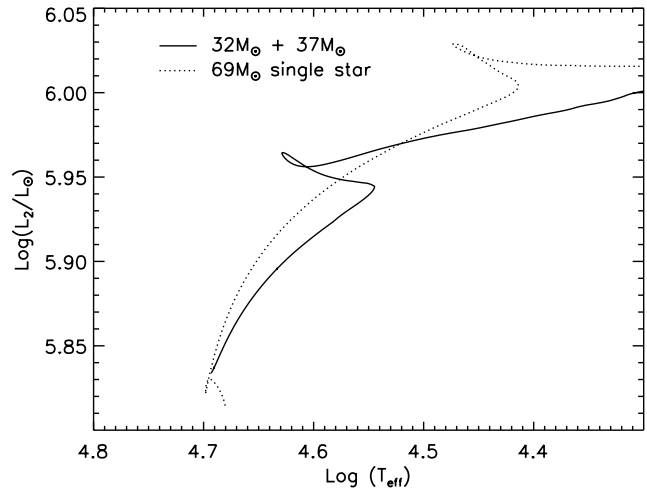}}
\end{center}
\caption{\textbf{Non-rejuvenation.} Hertzsprung-Russell diagram for a 32$\,$M$_\odot$ star that accreted  37$\,$M$_\odot$, and for a single star that starts its life with 69$\,$M$_\odot$. The stellar model that underwent mass accretion is the secondary component described in detail in the Letter. The luminosity does not include the correction due to tidal and rotational
distortions and the inclination of the system with respect to the line of sight.} 
\label{nonRejuvenation} 
\end{figure}

\begin{figure}
\begin{center}
\label{MassesNow}
\resizebox{10cm}{!}{\includegraphics{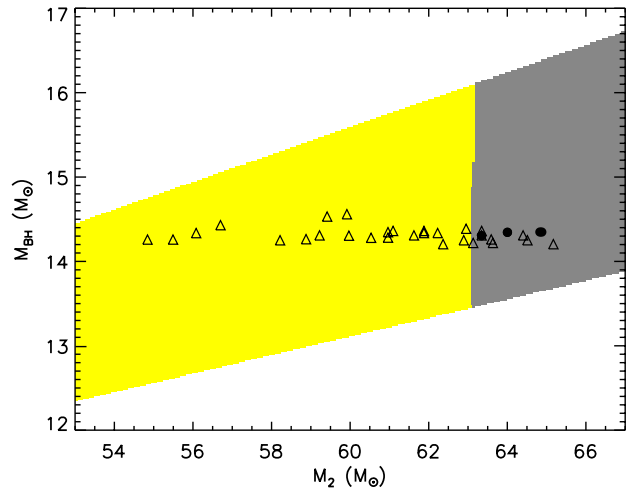}}
\end{center}
\caption{\textbf{Masses of the components at present.} The circles and triangles are the results of detailed binary star evolution calculations for all successful sequences for a distance to M33 of 840 $\pm$ 20$\,$kpc, and of 750-1017$\,$kpc, respectively. The grey and yellow shaded areas represent the observational constraints for a distance of 840 $\pm$ 20$\,$kpc, and of 750-1017$\,$kpc, respectively. The observational constraints are calculated given the dependence between the masses of the components $M\rm_{BH}$ = 6.19 + 0.13 $\cdot M\rm_{2}$, and accounting for uncertainties in the star's effective temperature, reddening, and apparent magnitude calculated through the ELC code. Some of the data points are omitted for clarity.} 
\label{MassesNow} 
\end{figure}
\begin{figure}
\begin{center}
\label{SNparams}
\resizebox{12cm}{!}{\includegraphics{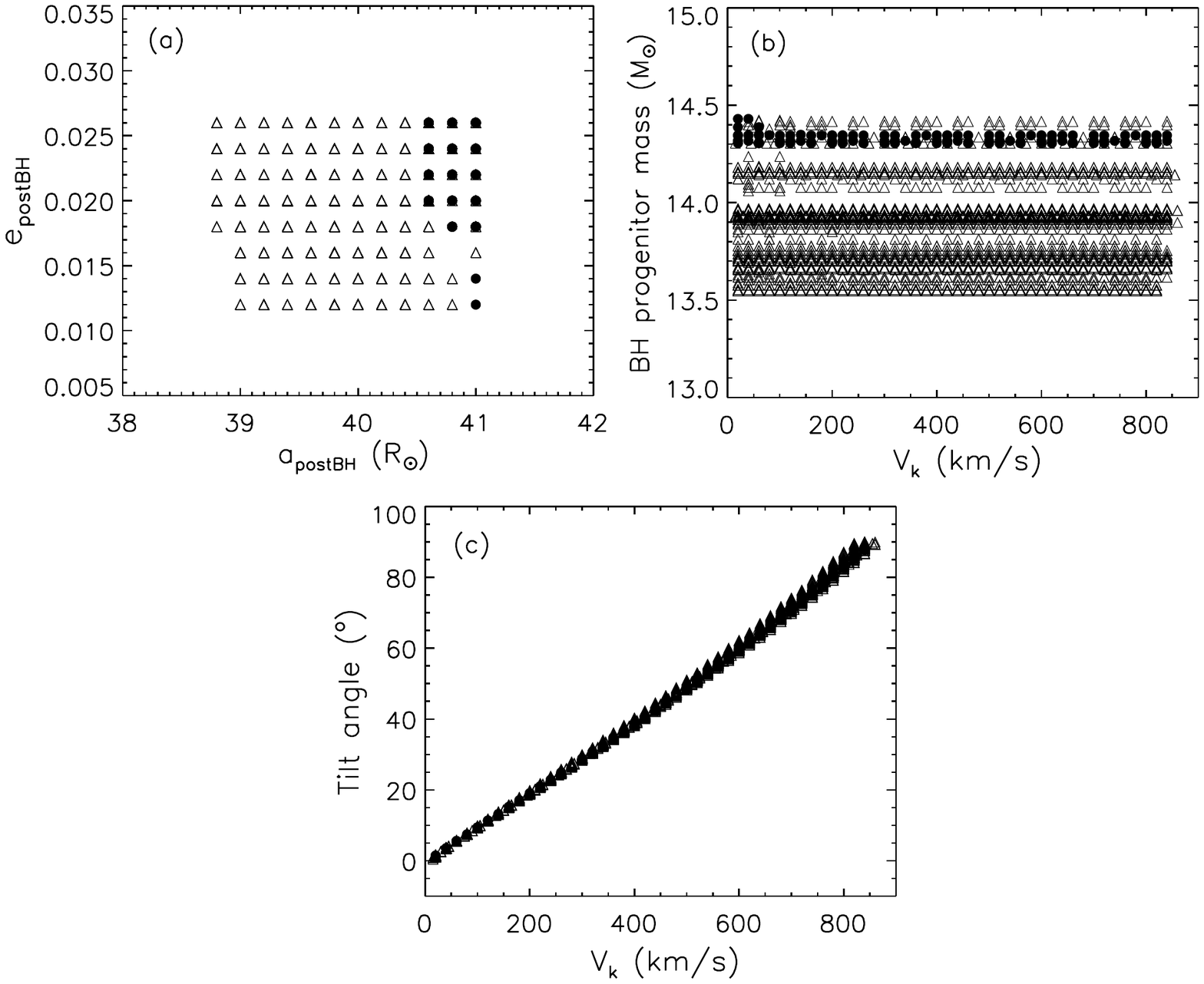}}
\end{center}
\caption{\textbf{Pre- and post- BH formation orbital parameters.} The circles and triangles are the results of detailed binary star evolution calculations for all successful sequences for a distance to M33 of 840 $\pm$ 20$\,$kpc, and of 750-1017$\,$kpc, respectively. (a) orbital eccentricity as a function of the orbital separation post-BH formation; (b) mass of the BH progenitor as a function of the kick magnitude; (c) change in the orbital inclination at BH formation as a function of the kick magnitude. The BH progenitor mass accounts for the 10\% of rest mass energy that is released as the BH's gravitational energy at collapse. According to our model, the BH progenitor mass lies within the observed range for the BH mass (15.65 $\pm$ 1.45$\,$M$_\odot$ for a distance of 840 $\pm$ 20$\,$kpc, and between 13.5-20$\,$M$_\odot$ for a distance of 750-1017$\,$kpc). Our model allows kicks from $\simeq10\,$km/s to $\simeq850\,$km/s. Some of the data points are omitted for clarity.} 
 \label{SNparams}
 \end{figure}